\documentclass[preprint,showpacs,preprintnumbers,amsmath,amssymb]{revtex4}
\usepackage{graphicx}
\usepackage{dcolumn}
\usepackage{bm}
\begin{document}
\title{Low carrier concentration crystals of the topological insulator Bi$_2$Te$_2$Se}
\author{Shuang Jia$^1$, Huiwen Ji$^1$, E. Climent-Pascual$^1$, M. K. Fuccillo$^1$, M. E. Charles$^1$, Jun Xiong$^2$, N. P. Ong$^2$ and R. J. Cava$^1$}
\affiliation{$^1$Department of Chemistry and $^2$Department of Physics, Princeton University, Princeton, NJ 08544, USA\\}

\begin{abstract}
We report the characterization of Bi$_2$Te$_2$Se crystals obtained by the modified Bridgman and Bridgman-Stockbarger crystal growth techniques. X-ray diffraction study confirms an ordered Se-Te distribution in the inner and outer chalcogen layers, respectively, with a small amount of mixing. The crystals displaying high resistivity ($> 1~ \mathrm{\Omega cm}$) and low carrier concentration ($\sim 5\times 10^{16}$/cm$^3$) at 4 K were found in the central region of the long Bridgman-Stockbarger crystal, which we attribute to very small differences in defect density along the length of the crystal rod. Analysis of the temperature dependent resistivities and Hall coefficients reveals the possible underlying origins of the donors and acceptors in this phase.

\end{abstract}


\maketitle


\section{Introduction}

The narrow-band-gap semiconductors (Bi,Sb)$_2$(Te,Se)$_3$ have been studied for over fifty years due to their use as thermoelectric materials \cite{handbookofthermo}.
Recently, these compounds have again come to the forefront in research in condensed matter physics, as the prototypical three-dimensional topological insulators (TIs), displaying gapped electronic bulk states and gap-less electronic surface states \cite{Fu_topo_2007, Fu_topo3D_2007, Moore_topo_2007, Hasan_topo_rev, Qi_topo_rev}.
The exotic, spin-locked Dirac metallic surface states of Bi$_2$Se$_3$ and Bi$_2$Te$_3$ have been revealed by angle-resolved photoemission spectroscopy (ARPES) studies \cite{Hsieh_BT_2009, xia_BS_2009, Chen_BT_2009}, as well as scanning tunneling microscopy (STM) experiments \cite{roushan_topological_2009}.
Investigating the charge transport characteristics of the surface states has, however, proven to be challenging, because the surface current is usually one or two orders of magnitude less than the bulk current \cite{Qu_BT_2010, butch_BS_2010}.
In binary or ternary narrow band gap semiconductors ($E_G<300~\mathrm{m eV}$) \cite{xia_BS_2009}, the small defect formation energies and the resultant significant defect densities often result in relatively large carrier densities $\sim 10^{18-19}$/cm$^3$, leading the bulk conductance to dominate over the surface conductance in transport probe measurements.
While previous materials research has primarily focused on heavily doped, low resistivity (Bi,Sb)$_2$(Te,Se)$_3$ for optimizing the thermoelectric figure of merit, the goal of the current research is to achieve highly resistive low bulk carrier concentration crystals that will facilitate the observation of the topological surface state transport.

The original studies of surface transport in TIs in the Bi$_2$X$_3$ family were focused on the binary Bi$_2$Se$_3$ and Bi$_2$Te$_3$ phases \cite{Qu_BT_2010, butch_BS_2010, Checkelsky_BS_2009}.
Near-stoichiometric Bi$_2$Se$_3$ crystals are always heavily-doped $n$-type materials due to the presence of charged electron-donating selenium vacancies \cite{handbookofthermo}, though a small amount of Ca substitution for Bi yielded $p$-type materials \cite{Hor_BS_2009, Checkelsky_BS_2009}.The defect equilibrium can be written as:
\begin{equation}
\mathrm{Bi_2Se_3} \rightleftharpoons 2\mathrm{Bi_{Bi}}+3\mathrm{V_{Se}^{\bullet \bullet }}+\frac{3}{2}\mathrm{Se_2(g)}+6e^{\prime}
\label{eqn-1}
\end{equation}
Near-stoichiometric Bi$_2$Te$_3$ crystals, on the other hand, are $p$-type due to the presence of anti-site defects, i.e. Bi on the Te sites \cite{handbookofthermo, Qu_BT_2010}, with the defect equilibrium:
\begin{equation}
\mathrm{Bi_2Te_3} \rightleftharpoons 2\mathrm{Bi_{Te}^{\prime}}+2h^{\bullet }+\mathrm{Te_2(g)}+\mathrm{Te_{Te}}
\label{eqn-2}
\end{equation}
Although small contributions of surface state conductance to the total conductivity have been reported in carefully grown, semiconductor-like Bi$_2$Te$_3$ single crystals, the resistivity of the binary compounds has not been reported to exceed 15 $\mathrm {m\Omega ~cm}$ \cite{Qu_BT_2010}; the reported surface conductance of Bi$_2$Te$_3$ contributes for those samples less than $0.3\%$ to the total conductance.

A transition from $p$-type to $n$-type behavior in the Bi$_2$(Te$_{1-x}$Se$_x$)$_3$ solid solution was reported 40 years ago \cite{seizo_BTS_1963, sokolov_BTS_2004}.
Very recent studies have revealed that when $x$ is close to $0.33$, yielding the structurally ordered phase Bi$_2$Te$_2$Se, crystals show high bulk resistivity at low temperatures, exceeding 1 $\mathrm {\Omega ~cm}$.
This, along with good topological surface state mobilities, results in a surface conductance contribution to the total conductivity of up to two orders of magnitude larger than is observed in Bi$_2$Te$_3$ \cite{ren_BTS_2010, xiong_BTS6ohm_2010}.
In some such Bi$_2$Te$_2$Se (BTS) crystals, clear Shubnikov-de Hass oscillations due to the surface states are observed \cite{ren_BTS_2010, xiong_BTS6ohm_2010}.
With its much larger surface current contribution, BTS may serve as an ideal model system for the study of spin-selected Dirac electron behavior, as well as a starting point for the future development of electronic devices based on topological surface states. This report describes how such crystals can be obtained.
 
The defects in the BTS ternary TI compound are anticipated to be more complex than those found in the binary systems, though the tendency toward Se site vacancies and Bi antisite defects, as seen in the binary compounds, is expected to be maintained; the effects of these and possible associated defects on the carrier density and mobility in the BTS ternary compound are still unclear. In this report, we describe the growth of the crystals of high quality BTS on which the successful surface transport measurements have been performed, focusing on the characterization of single crystal Bi$_2$Te$_2$Se obtained via two different crystal growth methods.

Powder X-ray diffraction (XRD) characterization of the crystals confirms that laboratory-grown Bi$_2$Te$_2$Se displays a fully ordered structure in which the Te and Se atoms occupy their own distinct crystallographic sites with the Te in the outer chalcogen layers and the Se in the inner chalcogen layer -- a factor that may be one of the underlying causes of the high surface mobility of the BTS phase.
Our electrical resistivity and Hall measurements reveal that the crystals' carrier concentration is highly sensitive to the Se site occupancy.
As expected for small band gap semiconductors, the crystals' quality from the perspective of the defects and the resultant carrier concentrations is strongly affected by small inhomogeneities in the chemical composition that occur during the crystal growth.
Realizing this difficulty, a Bridgman-Stockbarger method was employed to fine-tune the chemical composition of the crystals and optimize the carrier concentration.
We show this method to be a controllable way for synthesizing high resistance low carrier concentration bulk crystals of Bi$_2$Te$_2$Se.
The relation between the distribution of the donors and acceptors and the crystal properties is briefly discussed.

\section{Experiment}

Single crystals of Bi$_2$Te$_2$Se were prepared by two methods.
The first method, which we specify as a `modified Bridgman' method, is similar to that presented in ref. \cite{Hor_BS_2009, ren_BTS_2010}.
Five gram mixtures of high purity (5N) elemental Bi, Te and Se were sealed in quartz ampoules, and then heated up to 850~$^\circ $C for 1-2 days followed by cooling to 500 $^\circ $C at 6-12~$^\circ $C/h.
The samples were then annealed at 500~$^\circ $C for 3-4 days.
The crystals obtained by this method are within a large monolithic piece ($\sim 1\times 1\times 4~\mathrm{cm}^3$) that usually consists of approximately 10 grains presenting random crystal orientations.
For the powder XRD characterization of the laboratory-made BTS phase, the samples from the `modified Bridgman' method were in addition annealed at 400~$^\circ $C for over 2 weeks and then quenched in cold water. These samples were compared to samples that were directly quenched in water from the melt without any annealing to test for the robustness of the Te-Se crystallographic site ordering. Samples of the related compound Bi$_2$TeSe$_2$ were similarly prepared for comparison of the structures, and samples for which the Te/Se ratio was varied in small increments near the formula Bi$_2$Te$_2$Se for testing the effects of nonstoichiometry on the electrical properties were also prepared in this fashion. 
The samples for XRD were pulverized in liquid nitrogen and characterized by laboratory XRD using graphite monochromated $\mathrm{Cu}~K\alpha $ radiation (D8 Focus, Bruker) at room temperature.
Rietveld refinements were carried out using the FULLPROF program suite \cite{fullprof}.

The classic Bridgman-Stockbarger method was employed as the second crystal growth method. This method allows for fine-tuning of the chemical composition of the crystal near the stoichiometric Bi$_2$Te$_2$Se composition due to natural variations in stoichiometry along the directionally solidified crystal boule.
Thirty grams of mixture was sealed in a long internally carbon-coated quartz ampoule (20~$\mathrm{cm}$ long and 0.8~$\mathrm{cm}$ diameter).
This ampoule was tapered at the bottom in order to favor seed selection, and then placed in a vertical furnace.
The temperature profile of the furnace was set to ensure that the zone hotter than the melting temperature of the BTS was longer than the length of the liquid.
The temperature gradient at the furnace position crossing the melting point of the BTS was $\sim 30~^\circ $C/cm.
The ampoule was then lowered through the hot zone at the speed of 2-4~$\mathrm{mm/hour}$.
The crystal boules obtained, a characteristic one described in detail in this report, were about 14 $\mathrm{cm}$ long with fewer than 10 crystals, which were all grown with their $ab$ planes parallel to the long axis of the ampoule.
Such a uniform crystal morphology indicates that the boule is relatively homogeneous on a large scale, with its chemical composition gradually varying along the long axis during the directional solidification.
The boule analyzed was cut to seven pieces of about 2 $\mathrm{cm}$ equal length for investigating the electrical properties of different parts (see inset of Fig. \ref{fig5}).
The resistance and Hall measurements from 300~K to 10~K were performed in the $ab$ plane of the crystals in a Quantum Design Physical Property Measurement System (PPMS), and/or in a home-made resistance probe with a Keithley 2000 multimeter.
For the Hall effect measurements, the magnetic field was applied perpendicular to that plane.
The calculation of the carrier concertration and mobility is based on single band model for simplicity.
Considering the samples' expected variation in defect concentration, we selected multiple pieces from each region for measurements; only representative results are presented in this report.

\section{Results}

\begin{figure}
  \begin{center}
  \includegraphics[clip, width=0.7 \textwidth]{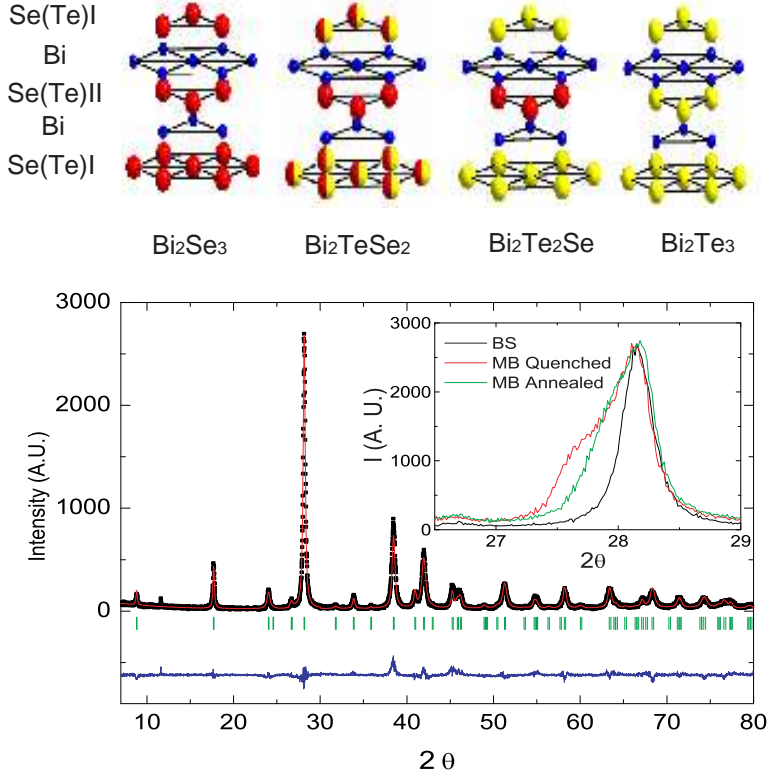}\\
  \caption{Upper panel: Schematic crystal structures of Bi$_2$Se$_3$, Bi$_2$TeSe$_2$, Bi$_2$Te$_2$Se and Bi$_2$Te$_3$. Lower panel: observed (open circles), calculated (solid line) and difference (lower solid line) XRD patterns of Bi$_2$Te$_2$Se grown via the Bridgman-Stockbarger method. Inset: comparison of the (105) peak of Bi$_2$Te$_2$Se grown via the Bridgman-Stockbarger method (black), and the modified Bridgman quenched (red) and annealed (green).}
  \label{Fig1}
  \end{center}
\end{figure}

\begin{table}
\caption{\label{table1} Powder XRD refinement results for a ground Bi$_2$Te$_2$Se crystal grown via the Bridgman-Stockbarger method. S.O.F.=site occupancy factor; U=thermal parameter (\AA $^2$), all constrained to be equal; $R_F=7.28$; $\chi ^2=2.33$.}
\begin{ruledtabular}
\begin{tabular}{lcccccc}
\multicolumn{3}{c}{$a=4.3067(1)$\AA} & \multicolumn{4}{c}{$c=30.0509(13)$\AA}\\
\hline
Atom & Wyck. & S.O.F. & x & y & z & U \\
\hline
Bi & 6c & 1.00       & 0 & 0 & 0.39681(8) & 0.0048(5)\\
Se$^{\mathrm{I}}$ & 6c & 0.043(7)  & 0 & 0 & 0.21155(8) & 0.0048(5)\\
Te$^{\mathrm{I}}$ & 6c & 0.957(7) & 0 & 0 & 0.21155(8) & 0.0048(5)\\
Se$^{\mathrm{II}}$ & 3a & 0.915(14)   & 0 & 0 & 0 & 0.0048(5)\\
Te$^{\mathrm{II}}$ & 3a & 0.085(14)  & 0 & 0 & 0 & 0.0048(5)\\
\end{tabular}
\end{ruledtabular}
\end{table}

Bi$_2$Se$_3$ and Bi$_2$Te$_3$ are isostructural five-layer tetradymite-type compounds, in which the layers stack in the sequence Se(Te)$^{\mathrm{I}}$-Bi-Se(Te)$^{\mathrm{II}}$-Bi-Se(Te)$^{\mathrm{I}}$ (Fig.\ref{Fig1}).
One crystallographic cell consists of three of these five-layer units stacked with rhombohedral symmetry, connected via weak van der Waals bonds between the Se(Te)$^{\mathrm{I}}$-Se(Te)$^{\mathrm{I}}$ layers in different units. The cleavage occurs between these van der Waals layers, with the exposed Se(Te)$^{\mathrm{I}}$ surface layer hosting the topological surface states.
The Se(Te) atoms in site $\mathrm{I}$ and site $\mathrm{II}$ are bonded to either three or six Bi nearest neighbors, respectively.
Such a difference in coordination of the two sites leads to different Se(Te)-Bi bondlengths: the Bi-Se(Te)$^{\mathrm{I}}$ bondlength is ~7\% shorter than the Bi-Se(Te)$^{\mathrm{II}}$ bondlength in both Bi$_2$Se$_3$ and Bi$_2$Te$_3$.
The longer bondlength of the central layer of Se(Te) indicates that Se(Te)$^{\mathrm{I}}$ can be considered as more ionic than Se(Te)$^{\mathrm{II}}$.
In order to investigate possible fully ordered ternary composition in the Bi$_2$(Te$_{1-x}$Se$_x$)$_3$ solid solution, Bi$_2$Te$_2$Se and Bi$_2$TeSe$_2$ were characterized by powder XRD (Fig. \ref{Fig1} and Table \ref{table1}).
Our refinements revealed that Bi$_2$Te$_2$Se has nearly fully ordered Te$^{\mathrm{I}}$ and Se$^{\mathrm{II}}$ layers, whereas Bi$_2$TeSe$_2$ has a fully disordered Se(Te)$^{\mathrm{I}}$ layer (a 50/50 random mixture of Se and Te) and a fully ordered Se$^{\mathrm{II}}$ layer (data not shown).
When the Te atoms are replaced by the more electronegative Se atoms in the Bi$_2$(Te$_{1-x}$Se$_x$)$_3$ solid solution, the Se atoms initially fill the central layer for $x\leq 1/3$ and then start to replace the Te$^{\mathrm{I}}$ atoms in the outside layer.
This result is consistent with early XRD experiments \cite{seizo_BTS_1963}.

Inspection of the XRD patterns of BTS samples grown by different methods reveals different peak shapes (Inset of Fig. \ref{Fig1}).
The diffraction peaks of the BTS samples grown by the modified Bridgman method are broad and asymmetric with a low angle shoulder.
We attribute the asymmetric broadening to a range of compositions Bi$_2$Te$_{2\pm \delta }$Se$_{1\mp \delta }$ in the crystals.
The XRD pattern of the quenched BTS shows significant double-peak character, indicating moderate phase separation, as is expected from fast-cooling a solid solution phase.
In comparison, the XRD pattern of BTS grown by the Bridgman-Stockbarger method has a much sharper and more symmetric peak shape, indicating that it has a more uniform composition.
This pattern was refined by using FULLPROF program based on a model with a 2:1 ratio of Te to Se, as free refinement of the composition did not show significant deviation from that formula.
As shown in Table \ref{table1}, the sample has near-perfect order of the Se and Te layers, with around 4\% disorder on the outer layers that support the topological surface states. We attribute the imperfect fit to some of the peaks to structural defects introduced during the grinding of the very soft BTS material. In contrast, Bi$_2$TeSe$_2$ shows full disorder on the outer layers, with 50/50 Te/Se occupancy.  
Being a nearly fully ordered crystal, Bi$_2$Te$_2$Se promises relatively low electron scattering that might be present due to general structural disorder at the surface, which is a Te layer, unlike what would be seen for Bi$_2$(Te,Se)$_3$ at higher selenium contents than Bi$_2$Te$_2$Se, where the surface layers would be substantially mixed.
Hence this work focuses on tuning the carrier concentration of Bi$_2$Te$_2$Se and measuring its general electronic transport properties to allow for its optimization for study of the transport due to the topological surface states.
As will be shown below, the relative homogeneity of BTS crystals grown via different methods has an impact on their observed electron transport properties.

\begin{figure}
  \begin{center}
  \includegraphics[clip, width=0.45 \textwidth]{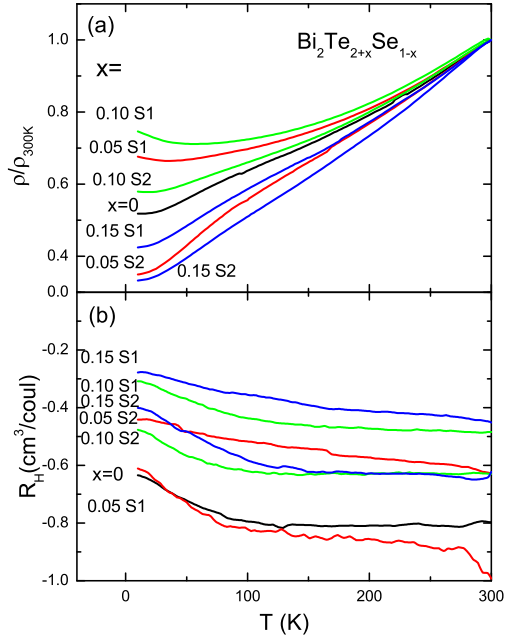}\\
  \caption{(a): The temperature dependent resistivities of Bi$_2$Te$_{2+x}$Se$_{1-x}$, S1 and S2 present different samples from the same batch. The resistivities for all the samples are $1-10~ \mathrm{m\Omega cm}$ at room temperature. (b): temperature dependent Hall coefficients of Bi$_2$Te$_{2+x}$Se$_{1-x}$ samples.}
  \label{Fig2}
  \end{center}
\end{figure}

As a starting point for the research, the temperature dependent resistance (shown as $\rho /\rho _{300\mathrm{K}}$) and Hall coefficient ($R_H$) were measured for stoichiometric Bi$_2$Te$_2$Se prepared by the modified Bridgman method.
As shown in Fig. \ref{Fig2}, our modified Bridgman growth Bi$_2$Te$_2$Se sample shows metallic behavior with a small negative $R_H$, leading to a large temperature-independent $n$-type carrier concentration $n=2.6\times 10^{19}$/cm$^3$ at 10~K.
Such a large carrier concentration and its trivial temperature-independent behavior indicate that the material is heavily doped by donors, resulting in a Fermi level ($E_F$) in the conduction band, far above the energy gap.
This $E_F$ position is similar to that seen in metallic Bi$_2$Se$_3$ in the ARPES measurements \cite{xia_BS_2009, Hor_BS_2009}. 

In order to explore the possibility of finding a low carrier concentration sample in the vicinity of the $p-n$ transition point in the Bi$_2$Te$_{2+x}$Se$_{1-x}$, solid solution, the series of samples synthesized near the stoichiometric formula Bi$_2$Te$_2$Se was characterized.
Because nominally stoichiometric BTS is $n$-type in the modified Bridgman crystal growths, samples of interest for reaching the crossover are expected to be rich in Te, where a greater tendency toward $p$-type behavior is expected (because Bi$_2$Te$_3$ is naturally $p$-type).
Figures \ref{Fig2} (a) and (b) show $\rho (T)$ and $R_H$ for crystals with $x=0.05$, $0.1$ and $0.15$ respectively.
None of them shows typical semiconducting $\rho (T)$ behavior, although two crystals from the $x=0.05$ and $0.10$ batches show weak negative temperature coefficients for $T<25$ K.
The $R_H$ for all the samples vary from $-1.0$ to $-0.2$ $\mathrm{cm}^3/ \mathrm{coul}$ at 10~K, leading to $n$-type carrier concentrations from $n=6\times 10^{18}$cm$^3$ to $3\times 10^{19}$cm$^3$.
The $p-n$ transition in Bi$_2$Te$_{2+x}$Se$_{1-x}$ was not found when $x\leq 0.15$.
This result is consistent with previously reported Seebeck coefficient measurements which showed a $p-n$ transition at $x=0.4$ \cite{sokolov_BTS_2004}.
Recently a high resistance sample of Bi$_2$Te$_{1.95}$Se$_{2.05}$ was reported \cite {ren_bsts_2011}; for our crystal growth methods all such compounds are metallic; this may be due the fact that majority carrier concentrations in this regime are strongly affected by very small amounts of non-stoichiometry, as discussed further below.

\begin{figure}
  \begin{center}
  \includegraphics[clip, width=0.45 \textwidth]{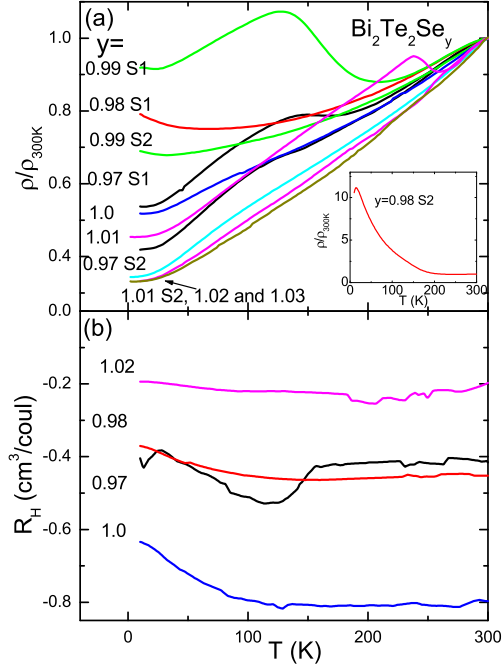}\\
  \caption{(a) Temperature dependent resistivities of Bi$_2$Te$_2$Se$_y$ samples. S1 and S2 are different samples from the same batch. The resistivities for all the samples are in the range $1-10~ \mathrm{m\Omega cm}$ at room temperature. (b)Temperature dependent Hall coefficients of Bi$_2$Te$_2$Se$_y$ samples.}
  \label{fig3}
  \end{center}
\end{figure}

Tuning the selenium concentration in the selenium excess or deficient Bi$_2$Te$_2$Se$_y$ series near $y=1$ can induce significant changes in the carrier concentration in BTS.
Figure \ref{fig3} shows the $\rho (T)$ and $R_H$ data for Bi$_2$Te$_2$Se$_y$ samples, where $y$ changes from $0.95$ to $1.02$.
For $y>1$, the values of $\rho _{10~\mathrm{K}}/\rho _{300~\mathrm{K}}$ are all smaller than what is seen in stoichiometric Bi$_2$Te$_2$Se.
The $R_H$ value for $y=1.02$ is $\sim -0.2$ cm$^3$/coul from 10~K to 300~K, leading to a strong metallic behavior.
This result indicates that simply adding Se to stoichiometric Bi$_2$Te$_2$Se crystal growths cannot compensate for the $n$-type majority carriers present at the stoichiometric ratio.
In contrast, most of the samples for $y<1$ show resistivities higher than the stoichiometric material.
The low temperature resistivities for these samples are still lower than their room temperature values, however, except for one sample from the batch $y=0.98$ (inset of Fig. \ref{fig3} (a)), which shows $\rho _{10\mathrm{K}}/\rho _{300\mathrm{K}}$ $\sim 10$.
It is worth noting that different crystal samples from one batch show dramatic differences in resistance when $y<1$.
Such significant sample-to-sample variation for crystals obtained from the `modified Bridgman' method presents a difficulty for their use in future research on surface state transport properties.

\begin{figure}
  \begin{center}
  \includegraphics[clip, width=0.45 \textwidth]{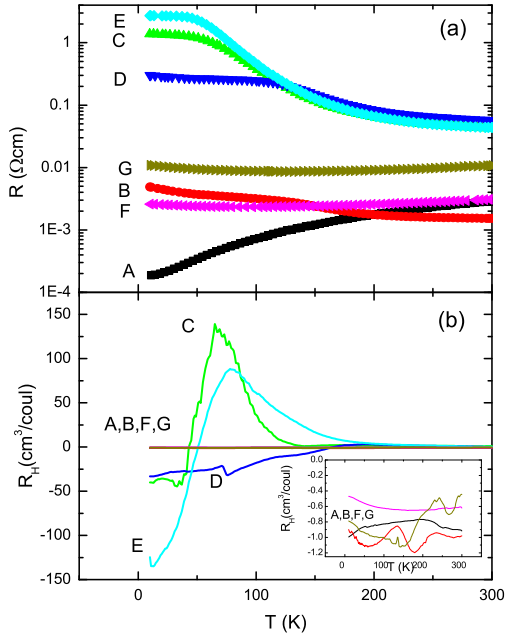}\\
  \caption{(a) Temperature dependent resistivities of Bi$_2$Te$_2$Se$_{.997}$ crystals grown by the Bridgman-Stockbarger method. (b) Temperature dependent Hall coefficients. Different parts of the crystal boule are designated alphabetically, with G the first-to-freeze section. Inset: blow-up of Hall coefficient for samples A, B, F and G.}
  \label{fig4}
  \end{center}
\end{figure}

\begin{figure}
  \begin{center}
  \includegraphics[clip, width=0.45 \textwidth]{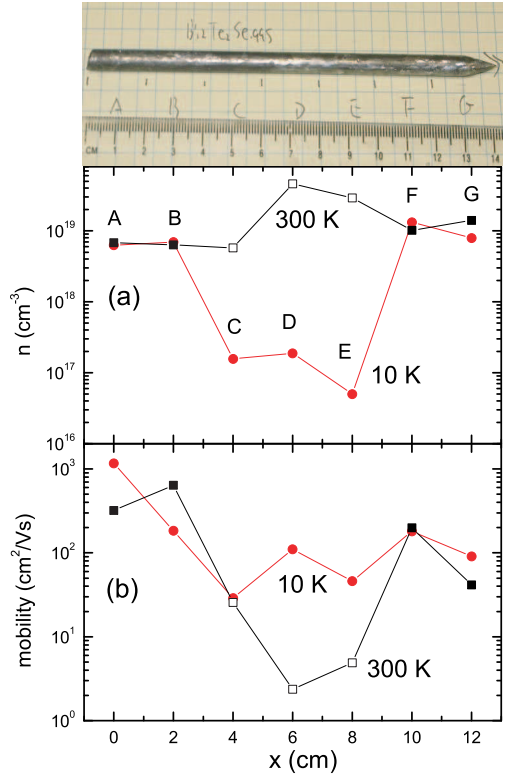}\\
  \caption{The carrier density (a) and mobility (b) at 10~K (circles) and 300~K (squares) of the Bi$_2$Te$_2$Se$_{.995}$ boule from the Bridgman-Stockbarger method with respect to the position along the boule. Samples C, D, and E are $p$ type at 300~K (open squares).}
  \label{fig5}
  \end{center}
\end{figure}

It is well known that a temperature gradient in a Bridgman growth furnace can induce a distribution of the chemical composition for a solid solution phase with a variation in melting point \cite{zonemelting}.
This mechanism was recently employed in the optical floating zone growth of the solid solution of La$_{2-2x}$Sr$_{1+2x}$Mn$_2$O$_7$ for exploring a fine structure of the charge and orbital phases in the diagram near $x=0.6$ \cite{zheng_charge_2008}.
Motivated by the partial success of the slightly Se-deficient crystals obtained in the modified Bridgman growths, a Bridgman-Stockbarger method was used for growing a crystal at a slightly Se deficient composition, Bi$_2$Te$_2$Se$_{.995}$.
Figure \ref{fig4} shows the $\rho (T)$ and $R_H$ data for different parts of the long Bridgman-Stockbarger crystal boule (see top of Fig. \ref{fig5}).
At the growth starting point (G and F), the samples show metallic resistivities and small negative Hall coefficients, which is similar to the metallic crystals obtained by the `modified Bridgman' method.
In the middle part of the boule (parts C, D, and E), the samples show semiconducting resistivity and much larger absolute values of $R_H$.
The three samples show positive $R_H$ between $0.2$ and $1$ $\mathrm{cm}^3/ \mathrm{coul}$ at 300~K, leading to hole concentrations of $p=0.5-3 \times 10^{18}$/cm$^3$.
$R_H$ significantly increases with decreasing $T$, leading to a positive maximum at about 70~K for samples C and E, and at about 200~K for sample D.
On cooling below the maximum, $R_H$ then decreases with decreasing $T$, crosses zero and shows a negative maximum at 10~K for all three samples, leading to the electron concentration $5\times 10^{16}/ \mathrm{cm}^3$--$2\times 10^{17}/ \mathrm{cm}^3$ at 10~K (Fig. \ref{fig5} (a)).
Given that the typical thickness of the samples is about $0.02$-$0.08$ $\mathrm{mm}$, this number is still one order of magnitude larger than the estimated surface electron contribution ($\sim 10^{12}$/cm$^2$), but the surface conductivity is similar in magnitude to the bulk conductivity due to the high surface mobility \cite{xiong_BTS6ohm_2010, ren_BTS_2010}.
The observed positive to negative transition of $R_H$ with temperature is similar to the previous reported experiments \cite{ren_BTS_2010, xiong_BTS6ohm_2010}, which clearly reveals the coexistence of two types of carriers in these BTS crystals.
The samples turn back to being metallic at the top part of the boule (parts A and B), in which the carrier concentration becomes large and negative $-1\times 10^{19} \mathrm{cm}^3$.
Although the sample shows metallic or semiconducting behavior at different positions along the boule, the properties are homogeneous on the scale of about 2~cm.
The Bridgman-Stockbarger crystal boules made with the starting compositions Bi$_2$Te$_2$Se$_{1.00}$ and Bi$_2$Te$_2$Se$_{.99}$ do not show semiconducting behavior (data not shown). This indicates that the Bridgman-Stockbarger method employed to grow this material induces less than a 1\% difference in chemical composition along the boule. We have shown here that the optimal defect densities can be obtained either accidentally in the crystals obtained by modified Bridgman growth or systematically at the $n$-to-$p$ crossover region in crystals obtained by the Bridgman-Stockbarger crystal growth method. 
The detailed transport properties of the surface states for the semiconducting crystals obtained here are reported elsewhere \cite{xiong_BTS_2011b, luo_BTS_2011}. 

The carrier concentrations ($n$) and mobilities ($\mu $) at 10~K and 300~K for the seven samples at 10~K are plotted in Fig. \ref{fig5}.
The mobility was calculated by assuming single majority carriers dominating at 10 and 300 K.
The metallic samples (A, B, F and G) show similar $n$ and $\mu $ values at 10~K and 300~K, which is expected in heavily doped semiconductors.
The semiconducting samples (C, D and E) show $\mu $ less than 100 $\mathrm{cm}^2\mathrm{/Vs}$ at 10~K, which is similar to previously reported bulk mobility values for BTS \cite{ren_BTS_2010, luo_BTS_2011}.
At 300~K, the mobility of the hole pocket is much less ($< 10~\mathrm{cm}^2\mathrm{/Vs}$).
Such small mobility may be due to the large effective hole mass.
The correlation between the mobility and carrier concentration for the semiconducting samples requires further investigation.

\section{Discussion and conclusion}

Based on our synthesis and characterization of a large number of BTS crystal samples, we can come to some conclusions about the relation between the chemical composition and the carrier distribution.
Stoichiometric BTS is always a heavy-doped $n$ type material showing metallic behavior, while slight tuning of the Te/Se ratio by changing the nominal composition at the percent level ($\leq 15$\%) does not lead to semiconducting samples.
In addition, starting with an excess amount of Se in the BTS growths, in an attempt to suppress the formation of Se vacancies, drives the materials more metallic.
A possible explanation for this is that the starting Bi$_2$Te$_2$Se$_{1+\delta}$ will result in a mixture of Bi$_2$(Te$_{2-\delta }$Se$_{ \delta}$)Se with excess Te, because the more electronegative Se atoms tend to bond with Bi atoms more strongly than they do with Te atoms.
Reducing the Se starting concentration on the other hand yields higher resistance samples, which are highly inhomogeneous when obtained by the modified Bridgman growth method.
This behavior indicates that reducing the Se starting concentration in BTS probably does not introduce the Se vacancies known to contribute $n$-type carriers in Bi$_2$Se$_3$.
We postulate instead that many of the Se vacancies that would have been the result of Se deficiency will be filled by Te atoms, while the excess Bi atoms that are present as a consequence will occupy the sites left vacant by the displaced Te, leading a $p$-type doping. We propose a more complex defect equilibrium for BTS:
\begin{equation}
\mathrm{Bi_2Te_2Se} \rightleftharpoons \mathrm{Bi_{Te}^{\prime}}+h^{\bullet }+(1-x)\mathrm{Te_{Se}^{\times }}+x\mathrm{V_{Se}^{\bullet \bullet }}+2xe^{\prime}+\frac{1}{2}\mathrm{Se_2(g)}+\mathrm{Bi_{Bi}}+\mathrm{Te_{Te}}+\frac{x}{2}\mathrm{Te_2(g)}
\label{eqn-3}
\end{equation}
This $n$-type carrier compensation mechanism is highly sensitive to the stating material ratio and the sample processing, leading to the high electronic inhomogeneity in the crystals grown by the modified Bridgman method. Further study will be required to determine whether our proposed defect model for BTS is correct. 

With its large temperature gradient at the freezing point, the classical Bridgman-Stockbarger method can introduce a continuous chemical composition distribution, and in the case of BTS can therefore allow for the achievement of highly resistive samples by selecting the appropriate position along the boule.
The semiconducting samples all manifest a majority carrier type changing from $p$-type to $n$-type on cooling from 300~K to 10~K.
A previous study suggested that the acceptors are from an impurity band with a gap $\sim 30~ \mathrm{meV}$ from the chemical potential \cite{ren_BTS_2010}, while our recent studies on BTS crystals under pressure reveal a satellite hole pocket $\sim 50~ \mathrm{meV}$ lower than the chemical potential \cite{luo_BTS_2011}.
Sample D shows an $R_H$ maximum at 200~K, higher in temperature than the current samples C and E and the samples reported in ref.\cite{ren_BTS_2010, luo_BTS_2011}, which indicates that its chemical potential may be closer to the bottom of the conduction band, leading to a larger gap with the hole pocket.
It is worth noting that the samples with higher temperatures for the $R_H$ maximum (such as sample D) show a larger $n$-type carrier concentration at 10~K.
We believe that the $n$-type carriers observed at low temperature in our measurements have a minor contribution from the surface metallic state and are mainly from a donor impurity band.
This impurity band is likely extended from states due to charged Se vacancies that are hybridized with the conduction band.
The chemical potential being closer to the conduction band in the band gap yields more donors at base temperature.

Our studies indicate that the BTS samples with small amounts of $p$-type doping made by the modified Bridgman or Bridgman-Stockbarger techniques are more likely to be semiconducting than the stoichiometric samples.
Given that the electrons at the bottom of the conduction band are lighter than the holes at the top of the valence band, the donors more easily form an impurity band near the electron pocket than the acceptors do near the hole pocket.
In other words, at equivalent n- and p-type defect concentrations, the Fermi level will be pinned near the isolated $p$-type energy levels in the band gap, yielding semiconducting samples, whereas the extended $n$-type impurity band leads Bi$_2$Te$_2$Se to be metallic for equivalent n-type carrier concentrations.
This assumption is consistent with our previous studies on Ca-doped Bi$_2$Se$_3$ \cite{Hor_BS_2009, Checkelsky_BS_2009}.
Although the bulk mobility of the semiconducting BTS samples is usually small $\sim 100~ \mathrm{cm}^2\mathrm{/Vs}$, the surface states manifest much higher mobility, leading to the observations of quantum oscillations previously reported. Further optimization of the crystal growth and doping of BTS is expected to yield excellent crystals for the study of topological surface state transport and the fabrication of surface-state-based electronic devices. 

\begin{acknowledgments}
The authors thank for the helpful discussion with N. Ni, M. Bremholm, J. Allred and K. Baroudi.
The crystallography work is supported by the NSF MRSEC program through DMR-0819860.
The crystal growth and characterization are supported by SPAWAR grant N6601-11-1-4110.
Substitutional chemistry work is supported by AFOSR FA9550-10-1-0533.
\end{acknowledgments}


\end{document}